\documentclass[10pt]{iopart}
\usepackage{graphicx,hyperref,amsfonts,amssymb}
\usepackage{epstopdf,color,bm,multirow,rotating,soul,comment}
\usepackage[caption=false]{subfig}
\usepackage{mathrsfs,amssymb}
\usepackage[absolute,overlay]{textpos}

\newcommand{\ket}[1]{|#1\rangle}

\begin{document}

\title[Generating pure single-photon states in a system of coupled microresonators]{Generating pure single-photon states via spontaneous four-wave mixing in a system of coupled microresonators}

\author{I.N. Chuprina$^{1,2}$, N.S. Perminov$^{1,3}$, D.Yu. Tarankova$^{4}$, and A.A. Kalachev$^{1,2,*}$}

\address{$^{1}$ Zavoisky Physical-Technical Institute, Kazan Scientific Center of the Russian Academy of Sciences, 10/7 Sibirsky Tract, Kazan 420029, Russia}
\address{$^{2}$ Kazan Federal University, 18 Kremlyovskaya Str., Kazan 420008, Russia}
\address{$^{3}$ Kazan Quantum Center, Kazan National Research Technical University n.a. A.N.Tupolev-KAI, 10 K. Marx, Kazan 420111, Russia}
\address{$^{4}$ Institute of Radio-Electronics and Telecommunications, Kazan National Research Technical University n.a. A.N.Tupolev-KAI, 10 K. Marx, Kazan 420111, Russia}
\ead{a.a.kalachev@mail.ru}

\vspace{10pt}
\begin{indented}
\item[]\today
\end{indented}

\begin{abstract}
We present the optimal design for an on-chip single-photon source based on spontaneous four-wave mixing in a system of coupled ring microresonators, which provides frequency uncorrelated joint spectral amplitude of the biphoton field and thereby generation of pure single-photon heralded states. A simple method is proposed for suppressing negative dispersion effects by optimizing the controlled spectroscopic parameters of the system. It shown that the optimal coupling parameters, in combination with the optimal spectral width of the pump pulse, give rise to the highest purity of the heralded photons for a given pump linewidth.
\end{abstract}

\pacs{42.50.Dv, 42.65.Lm, 42.65.Wi, 42.65.Yj}
\vspace{2pc}
\noindent{\it Keywords}: single-photon source, microresonator, spontaneous four-wave mixing, spectrum optimization, integrated optics devices, quantum information and processing
\maketitle
\ioptwocol

\section{Introduction}
Developing single-photon sources is an important task of optical quantum technologies \cite{Eisaman:2011cc,Takeuchi2014,Caspani:2017gq}. In particular, CMOS-compatible on-chip devices are especially demanded for creating scalable and compact quantum photonic circuits \cite{Politi2009,Harris:2014kj,Caspani:2017gq}. In this respect, heralded single-photon sources based on spontaneous four-wave mixing (SFWM) in microring resonators are of great interest since they allow one to achieve within the framework of integrated optics high efficiency of the nonlinear process \cite{Gaeta:2008bc,Azzini:2012io,Savanier:2016kb}, narrow spectral width of generated photons \cite{Reimer:2014ev}, and to approach deterministic emission of them using multiplexing techniques \cite{Collins:2013eu,Heuck:2018km}. The latter is expected to be quite efficient when using photon number resolving detectors \cite{Christ:2012bs}. In addition, the sources can be designed to produce pure single-photon states (transform--limited single-photon wave packets) \cite{Helt:10,Vernon:17}, which is crucially important for observing quantum interference effects and implementing optical quantum computing \cite{Kok2016}. It is also worth noting that cryogenic temperatures in this case are not required, in contrast to single-photon sources based on single quantum emitters such as quantum dots or color centers.

In the present paper, we develop a scheme for generating pure single-photon states via SFWM in a system of coupled ring microresonators. In the SFWM process, two pump laser photons are converted into a pair of daughter photons, usually called signal and idler, in a third-order nonlinear optical material. The photon number correlation between the resulting fields can be exploited to herald the existence of one photon by detection of its partner, which underlies the conditional preparation of single-photon states. Energy conservation requires the signal and idler photons to be generated at frequencies that are symmetrically distributed around the pump frequency. In a general case, due to such a spectral correlation, the heralded photons prove to be in a mixed state. High purity of the emitted photons is achieved when the joint spectral amplitude (JSA) of the biphoton field is a factorable function in the frequency domain \cite{Ren2005}, which is possible for a sufficiently broadband pump field. Similar to \cite{Vernon:17}, for the latter to be used we take advantage of a smaller pump quality factor, which makes the linewidth of the microresonator for the pump broader than those for the signal and idler fields. However, instead of using two coupling points via Mach--Zehnder interferometers, we suggest to use additional microrings. An important advantage of our scheme is that the microrings can be fabricated of special sizes and tuned in resonance only with three interacting modes thereby making additional spectral filtering unnecessary. In addition, the present scheme may be easily realized not only with microring resonators but also with other types of resonators such as microspheres and microtoroids.

\section{Basic model}
\begin{figure}[t]
	\includegraphics[width = 0.45\textwidth]{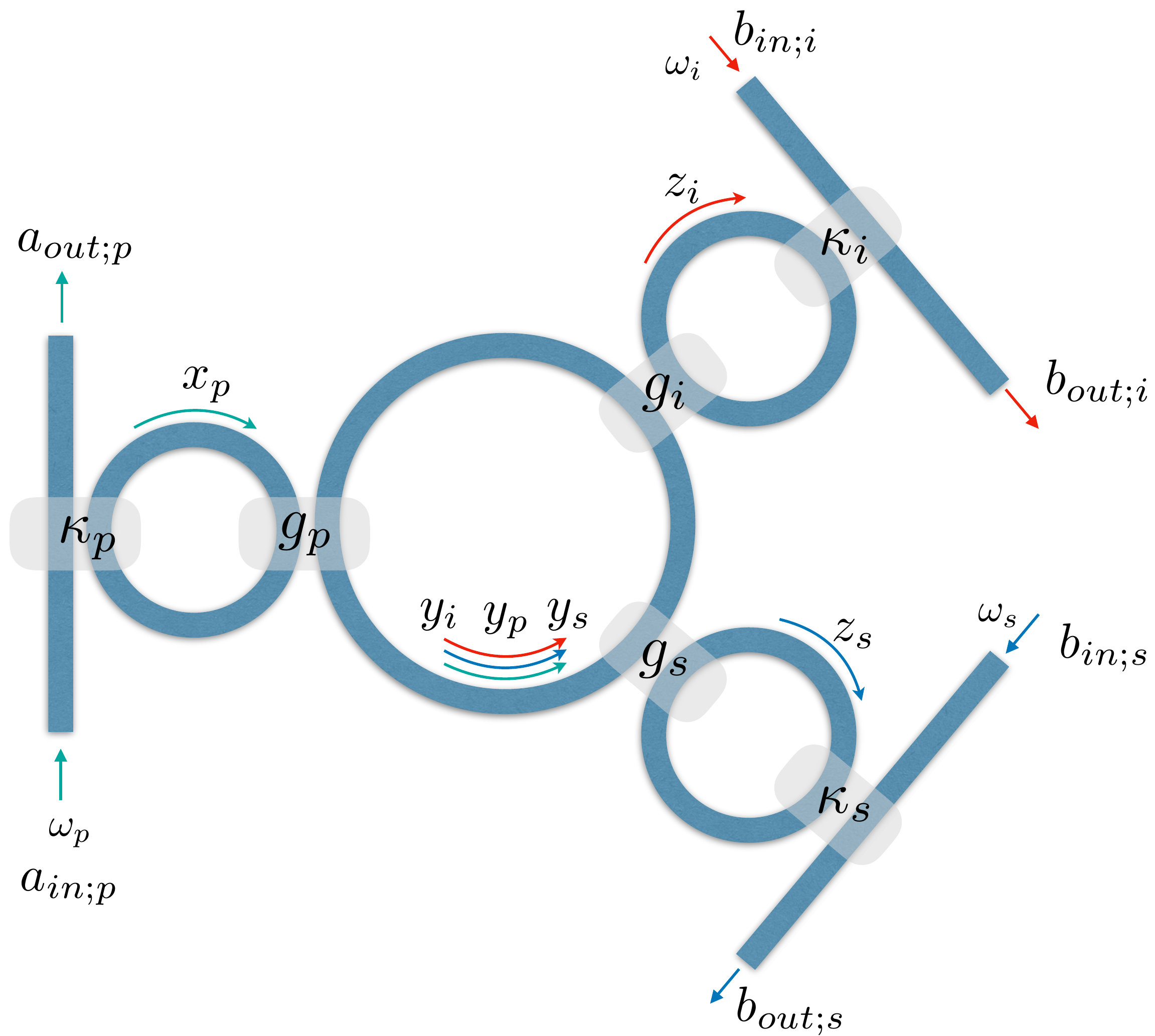}
	\caption{Scheme of coupled microring resonators connected with three waveguides (buses).}
	\label{Scheme}
\end{figure}

In the present paper, we consider a system of three microring resonators coupled to a central one and connected with strait waveguides (buses)  (Fig.~\ref{Scheme}). 
The SFWM process occurs in the central ring, while other rings are used for loading the pump field and unloading the generated photons. For simplicity, we consider a degenerate pump scheme. It  is  assumed  that  the  pump  field  corresponds  to  a  resonator mode in the zero-dispersion region of the central ring microresonator so that the signal and idler photons can be emitted into the adjacent modes that are separated from the pump mode by equal frequency intervals (the group velocity dispersion is negligible). One of the outer rings is tuned in such a way that one of its resonances coincides with the pump mode of the central ring, whereas other resonances do not coincide with the signal and idler modes. On the contrary, other outer rings should be out of resonance with the pump mode but in resonance with the signal and idler modes. When the free spectral range of the outer rings is two times smaller than that of the central ring, the system is similar to that of \cite{Vernon:17}. However, it is possible to make these rings of other sizes so that only three modes of the central ring prove to be effectively coupled to the strait waveguides.

The Hamiltonian for the system is
\begin{equation}\label{Hamilt}
\mathcal{H} = \mathcal{H}_{sys}+\mathcal{H}_{bath}+\mathcal{H}^{internal}_{int}+\mathcal{H}^{bath}_{int},
\end{equation}
where
\begin{eqnarray*}\label{exten_hamilt2}
\mathcal{H}_{sys} = &&\hbar\omega_{0x,p}\, x_p^\dagger x_p + \sum_{n=p,i,s} \hbar\omega_{0y,n}\,y_n^\dagger y_n \\
&&+ \sum_{m={i,s}} \hbar\omega_{0z,m}\, z_m^\dagger z_m
\end{eqnarray*}
is the free-field Hamiltonian for the cavities,
\begin{equation*}
\mathcal{H}_{bath} = \int d\omega\,\hbar\omega \Big[ a_p^\dagger (\omega) a_p(\omega) + \sum_{m=i,s} b_m^\dagger(\omega) b_m(\omega) \Big]
\end{equation*}
is the external bath Hamiltonian,
\begin{equation*}
\mathcal{H}_{int}^{internal} = i\hbar g_p x_p^\dagger y_p+i \hbar g_i z_i^\dagger y_i + i \hbar g_s z_s^\dagger y_s+ h.c.
\end{equation*}
is the coupling between the rings, and
\begin{eqnarray*}
\mathcal{H}_{int}^{bath} = &&\frac{i\hbar}{\sqrt{2\pi}} \int d\omega\, \big[ \sqrt{\kappa_p}\, x_p^\dagger a_p(\omega) +
\sqrt{\kappa_i}\,z_i^\dagger b_i(\omega)\\
&&+\sqrt{\kappa_s}\,z_s^\dagger b_s(\omega)+h.c.\big]
\end{eqnarray*}
is the coupling to the external modes. Here $m =\{i,s\}$, $n=\{i,s,p\}$, $\omega_{0x,p},\omega_{0y,n},\omega_{0z,m}$ are the central frequencies of the microresonators, $\kappa_p$, $\kappa_{i,s}$ are the coupling parameters between the waveguide and rings, $g_p$, $g_{i,s}$ are coupling parameters between the rings, $x_p$, $y_n$ and $z_m$ are annihilation operators for the photons corresponding to the different modes in the rings.
Non-zero commutation relations read: $[x_p,x_p^\dagger]=[y_n,y_n^\dagger]=[z_m,z_m^\dagger]=1$ and $[a_p(\omega), a_p^\dagger(\omega')]=[b_m(\omega),  b_m^\dagger(\omega')]=\delta(\omega-\omega')$. 

By applying the input--output formalism \cite{Walls}, from Eq.~(\ref{Hamilt}) we obtain the following Heisenberg-Langevin equations:
\begin{equation}\label{Heis_Lang1}
\eqalign{
\left[\partial_t+i\omega_{0x,p} +\frac{\kappa_p}{2}\right] x_p -g_p y_p= \sqrt{\kappa_p}\, a_{in;p},\cr
[\partial_t+i\omega_{0y,i}] y_i+g_i z_i =0,\cr 
[\partial_t+i\omega_{0y,p}] y_p+g_p x_p=0,\cr
[\partial_t+i\omega_{0y,s}] y_s+g_s z_s=0,\cr
[\partial_t+i\omega_{0z,i} +\frac{\kappa_i}{2}] z_i -g_i y_i=\sqrt{\kappa_i}\,b_{in;i},\cr
[\partial_t+i\omega_{0z,s} +\frac{\kappa_s}{2}] z_s -g_s y_s=\sqrt{\kappa_s}\,b_{in;s,}\cr
a_{in;p}-a_{out;p} = \sqrt{\kappa_p}\, x_p,\cr
b_{in;i}-b_{out;i}=\sqrt{\kappa_i}\, z_i,\cr
b_{in;s}-b_{out;s}=\sqrt{\kappa_s}\, z_s,
}
\end{equation}
where we used Markov approximation $\kappa_p (\omega) \approx \kappa_p = \rm{const}$, $\kappa_{i}(\omega) = \kappa_{s}(\omega) \approx \kappa_{is}= \rm{const}$, $g_p(\omega) \approx g_p = \rm{const} $ and $g_i(\omega) = g_s (\omega)  \approx g_{is} = \rm{const}$. In what follows, we also assume that the same modes in the different rings are matched with each other: $\omega_{0x,p}=\omega_{0y,p}=\omega_{0p}$, $\omega_{0z,m}=\omega_{0y,m}=\omega_{0m}$, which is a natural condition for an efficient energy transfer.

\section{Input--output relations}	
By taking the Fourier transform of (\ref{Heis_Lang1}), we obtain the system of algebraic equations 
\begin{equation}\label{Heis_Lang_fourier}
\eqalign{
\left[i\Delta_{p} +\frac{\kappa_p}{2}\right] x_p(\omega) -g_p y_p(\omega)= \sqrt{\kappa_p}\, a_{in;p}(\omega),\cr
i\Delta_{p} y_p(\omega)+g_p x_p(\omega)=0,\cr
i\Delta_{m} y_m(\omega)+g_{is} z_m(\omega)=0,\cr
\left[i\Delta_{m} +\frac{\kappa_{is}}{2}\right] z_m(\omega) -g_{is} y_m(\omega)=\sqrt{\kappa_{is}}\,b_{in;m}(\omega),\cr
a_{in;p}(\omega)-a_{out;p}(\omega) = \sqrt{\kappa_p}\,x_p(\omega),\cr
b_{in;m}(\omega)-b_{out;m}(\omega)=\sqrt{\kappa_{is}}\, z_m(\omega),
}
\end{equation}
where we introduce $\Delta_{n} = \omega_{0n}-\omega$, and for all the annihilation operators the Fourier transform is defined as $u(t)=\frac{1}{\sqrt{2\pi}}\int d\omega\, e^{-i\omega t} u(\omega)$. 

Let us consider the case when $a_{in;p}=1$, and $b_{in;m}=0$, which corresponds to the loading of the pump field. Then we obtain the input-output relations for the pump field operators
\begin{equation}\label{in-out_pump}
y_p(\omega)=M_p a_{in;p}(\omega)=\frac{2g_p\sqrt{\kappa_p}\,a_{in;p}(\omega)}{-2\Delta_{p}^2+2g_p^2+i\Delta_{p}\kappa_p}.
\end{equation}
Similarly, in the case when $a_{in;p}=0$, and $b_{in;m}=1$, which corresponds to the loading of the signal and idler fields (and unloading them for the reversed time), we get
\begin{equation}\label{in-out_m}
y_m(\omega)=M_m b_{in;m}(\omega)=\frac{2g_{is}\sqrt{\kappa_{is}}\,b_{in;m}(\omega)}{-2\Delta_{m}^2+2g_{is}^2+i\Delta_{m}\kappa_{is}}.
\end{equation}
To express the cavity field operators $y_n$ in terms of the output fields $a_{out;p}(\omega)$, and $b_{out;m}(\omega)$, $M_n$ in Eqs.~(\ref{in-out_pump}), (\ref{in-out_m}) should be replaced by $M^*_n$.

\section{Optimal coupling}
To suppress the phase dispersion in the central microring, which is necessary for the effective loading of the pump field at frequency $\omega_{0p}$ into it and unloading the generated photons at the frequencies $\omega_{0i}$, $\omega_{0s}$ from it through the outer microrings, we apply the following condition of the closeness of the frequency dependence of the phase to the linear one:
\begin{equation}
\Bigl. \partial^l_{\omega}\textrm{Argument}(M_n)\Bigr|_{\omega=\omega_{0n}}=0,\quad l=2,3,...
\end{equation}
In our system, we can impose the condition for $l=3$, which leads to
\begin{equation}
\eqalign{
\Bigl. \partial^3_{\omega}\textrm{Arctan}\left[\frac{\Delta_{p}\kappa_p}{2\Delta_{p}^2-2g_p^2}\right] \Bigr|_{\omega=\omega_{0p}}=0,\cr
\Bigl. \partial^3_{\omega}\textrm{Arctan}\left[\frac{\Delta_{m}\kappa_{is}}{2\Delta_{m}^2-2g_{is}^2}\right] \Bigr|_{\omega=\omega_{0m}}=0,
}
\end{equation}
and obtain the following optimal ratios between the coupling parameters
\begin{equation}\label{optimalratio}
g_{p,\rm{opt}}=\kappa_p/\sqrt{12},~ g_{is,\rm{opt}}=\kappa_{is}/\sqrt{12}.
\end{equation}
The remaining conditions for $l>3$ lead to $\kappa_{p,is}\rightarrow\infty$. Physically, this means that the maximum possible experimental values of $\kappa_{p,is}$ should be used.

\begin{figure}[t]
	\includegraphics[width=0.45\textwidth]{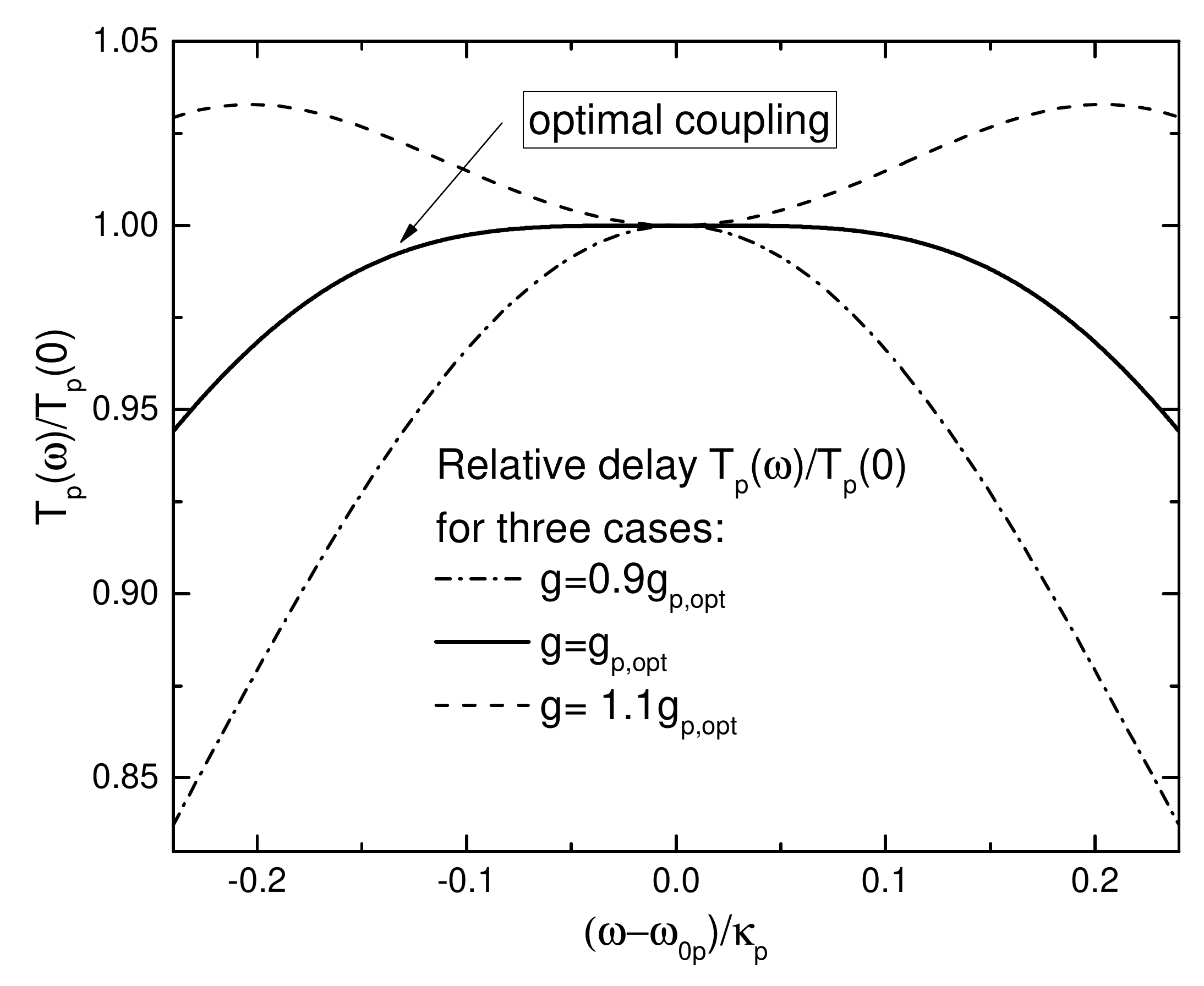}
	\caption{Relative delay $T_p(\omega)/T_p(0)$ as a function of $(\omega-\omega_{0p})/k_p$ for three cases: $g_p=0.9g_{p,\rm{opt}}$ (dashed line), $g_p=g_{p,\rm{opt}}$ (solid line), and $g_p=1.1g_{p,\rm{opt}}$ (dash-dot line), where $g_{p,\rm{opt}}=\kappa_p/\sqrt{12}$.}
	\label{opt_coupl}
\end{figure}

To conveniently visualize the dispersion effects, we introduce the delay function $T_p(\omega)=\textrm{Argument}(M_p)/(\omega-\omega_{0p})$ which shows the difference in a time delay of signals at different frequencies near the central frequency $\omega_{0p}$ (the case of unloading of signal and idler fields can be described similarly). Fig.~\ref{opt_coupl} demonstrates the difference between the three cases, $g_p = \{0.9g_{p,\rm{opt}}, g_{p,\rm{opt}}
, 1.1g_{p,\rm{opt}}\}$.
It can be seen that the maximum size of the plateau, corresponding to the maximum suppression of negative dispersion effects for the pump field in the central ring, is attained for the ratio (\ref{optimalratio}). Such dispersion suppression is necessary to improve the quality of the heralded photons to the extent possible.

\section{Purity of the state}
The SFWM theory in the microrings was developed in a number of papers \cite{savanier2016photon, Helt:10, Chen:11, Helt:12, Camacho:12, PhysRevA.92.033840, PhysRevA.91.053802, Vernon:16, Vernon:17}. To calculate the state of the biphoton field, we can take advantage of the first-order perturbation SFWM theory in an optical waveguide or fiber (see, for example, Ref. \cite{GarayPalmett:2010hf}), which is modified taking into account the input--output relations. This approach was used for the analysis of the cavity-assisted SFWM in Refs.~\cite{Chen:11,GarayPalmett:2012bv,Chuprina2017}.

The SFWM process in the central resonator is described by the effective Hamiltonian
\begin{equation}
\mathcal{H}_{SFWM}(t) = \zeta y_p(t) y_p(t) y^\dagger_s(t) y^\dagger_i(t),
\end{equation}
where $\zeta$ is the effective nonlinearity that takes into account $\chi^{(3)}$ of the nonlinear material, microresonator mode functions and other parameters, which are not important for the present analysis.

By applying the first-order perturbation theory, the state vector of the generated biphoton field is calculated as $\ket{\psi}=[1-i/\hbar \int dt\, \mathcal{H}_{SFWM}(t)]\ket{0}\ket{\alpha}$, which gives
\begin{eqnarray}
\ket{\psi}=&&\ket{0}\ket{\alpha}-\frac{i\zeta}{\hbar(2\pi)^2} \int dt d\omega_p d\omega_i d\omega_s   \times\nonumber\\
&& y_p(\omega_p) y_p(\omega_p)y^\dagger_i(\omega_i) y^\dagger_s(\omega_s) e^{i\Delta\omega t} \ket{0}\ket{\alpha},
\end{eqnarray}
where $\ket{0}=\ket{0_s}\ket{0_i}$ is the vacuum state of the signal and idler fields, $\ket{\alpha}$ is the coherent state of the pump field  with a complex amplitude $\alpha$ (i.e., $y_p(t)\ket{\alpha}=\alpha(t)\ket{\alpha}$), and $\Delta\omega=2\omega_p-\omega_i-\omega_s$ is the frequency detuning. 
Now, taking into account the input--output relations (\ref{in-out_pump},\ref{in-out_m}) we obtain
\begin{eqnarray}
\ket{\psi} = &&\ket{0}\ket{\alpha}-
\frac{i\zeta}{\hbar \sqrt{2\pi}^3} \int  d\omega_i d\omega_s  \mathcal{F}(\omega_i, \omega_s) \times\nonumber\\ && y^\dagger_{out;i}(\omega_i) y^\dagger_{out;s}(\omega_s) \ket{0}\ket{\alpha},
\end{eqnarray}
where 
\begin{equation}
\mathcal{F}(\omega_i, \omega_s)=\mathcal{I}_p (\omega_i,\omega_s) M_i(\omega_i) M_s(\omega_s)
\end{equation}
is the JSA of the biphoton field, and
\begin{eqnarray}
\mathcal{I}_p (\omega_i,\omega_s) =
&&\int d\omega_p M_p(\omega_s+\omega_i-\omega_p)M_p(\omega_p) \times\nonumber\\
&& \alpha(\omega_s+\omega_i-\omega_p) \alpha(\omega_p)
\end{eqnarray}
is the convolution of the spectral amplitude of the pump field $\alpha(\omega_p)$ in the resonator. 

To illustrate spectral correlations between the emitted photons, it is convenient to use the joint spectral intensity (JSI) $\mathcal{P}(\omega_i, \omega_s)=\left|\mathcal{F}(\omega_i, \omega_s)\right|^2$.
In addition, for quantitative analysis we can take advantage of a Schmidt decomposition of the JSA \cite{Law:2000fm,Fedorov:2014dc}, which can be written as
\begin{equation}
\mathcal{F}(\omega_i, \omega_s)=\sum_n\sqrt{\lambda_n}\,\psi_n(\omega_i)\phi_n(\omega_s),
\end{equation}
where the Schmidt coefficients satisfy the condition $\sum_n\lambda_n=1$. Then the Schmidt number $K=1/\sum_n\lambda_n^2$ is usually used as a measure of entanglement in the photon pairs \cite{Law:2004hw,Fedorov:2006jk}, while the purity of the heralded single-photon state is equal to $\gamma=1/K$. A two-photon state for which $K=1$ (the minimum value) represents a factorable state, which exhibits no spectral entanglement and gives rise to pure heralded single photons.
\begin{figure}[ht]
	{
\includegraphics[clip,width=\columnwidth]{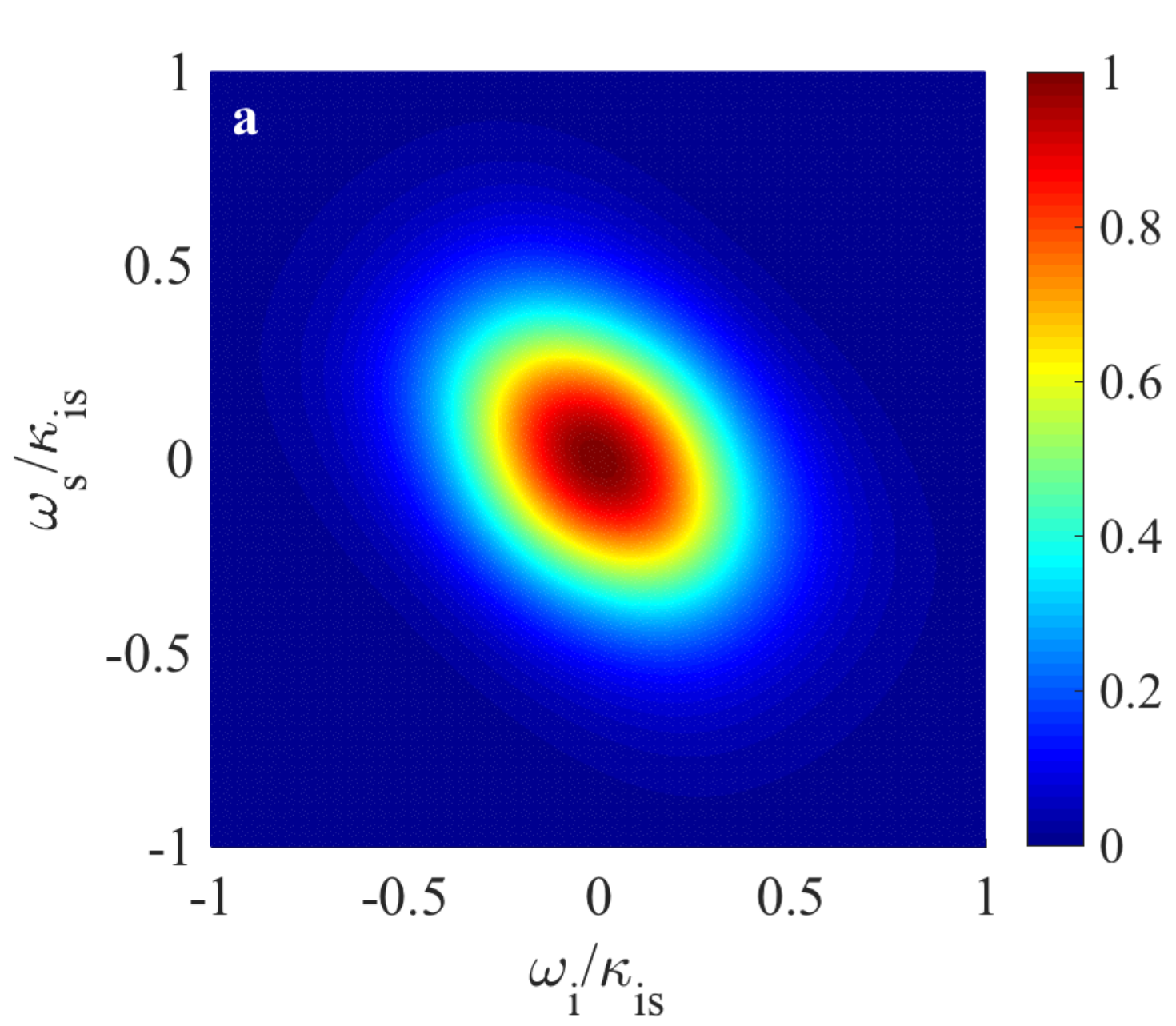}
    }   
	{\includegraphics[clip,width=\columnwidth]{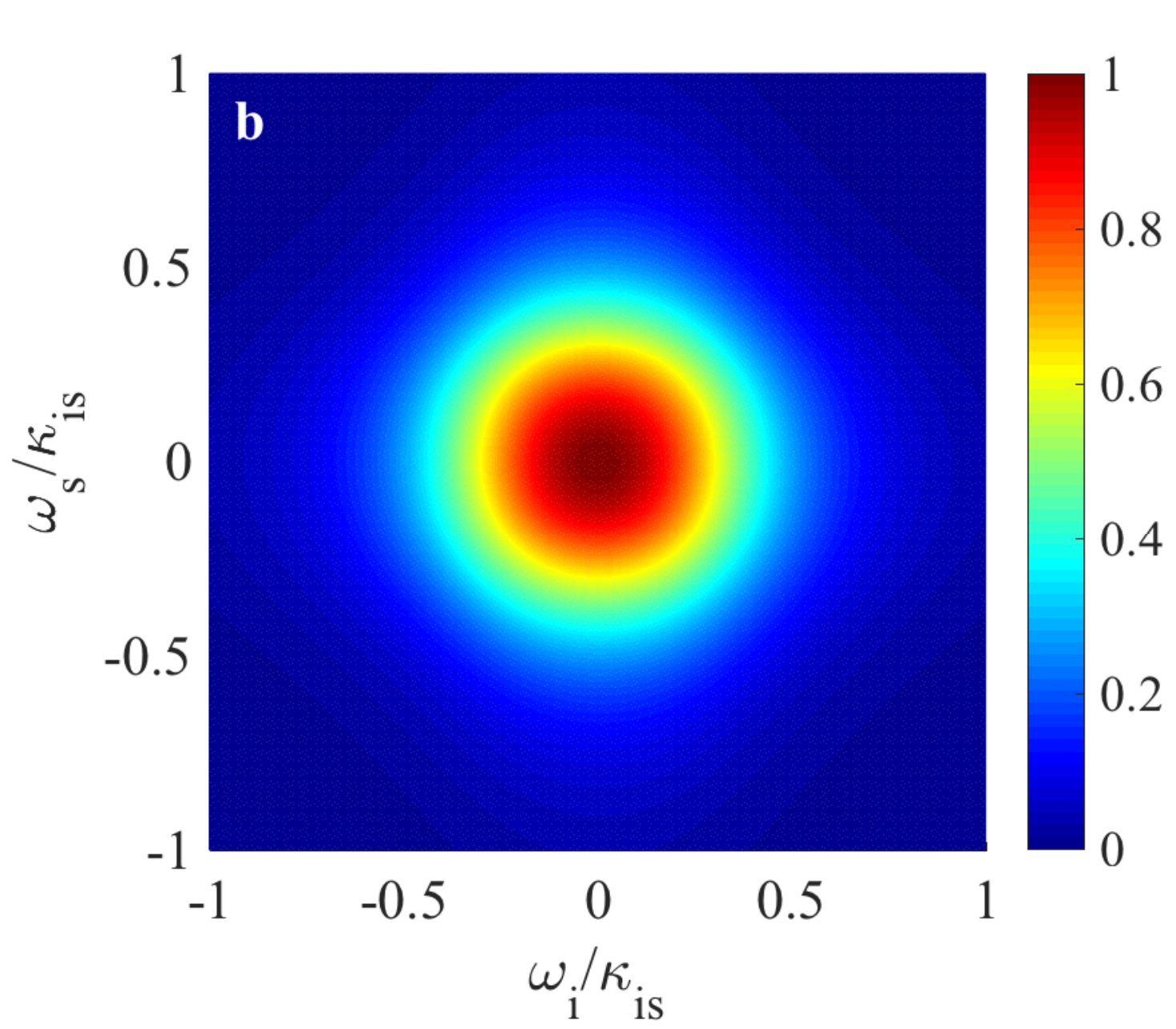}
    }
	\caption{JSI of the biphoton field for the equal resonator linewidths $\kappa_p=\kappa_{is}$ (a) and for the broader pump linewidth $\kappa_p=10\kappa_{is}$ (b). The Schmidt number is $K = 1.07$ $(\gamma = 0.94)$ and $K = 1.00006$ $(\gamma = 0.9999)$, respectively.}
	\label{JSI}
\end{figure}
Fig.~(\ref{JSI}) illustrates the JSI distributions calculated numerically for the equal resonator linewidths $\kappa_p=\kappa_{is}$ and for the broader pump linewidth $\kappa_p\gg\kappa_{is}$.
In both cases, the optimal ratio between the coupling parameters (\ref{optimalratio}) is maintained, and the pump pulse is assumed to be Gaussian, $\alpha(\omega)=(2\pi\sigma)^{-1/4}\exp(-(\omega-\omega_{0p})^2/4\sigma)$, with a spectral width, $\Delta\omega_{1/2}=\sqrt{8\sigma\ln 2}$, optimized for providing the minimum value of the Schmidt number. Similar to \cite{Vernon:17}, the calculations show near perfect separability of the biphoton field in the case of the broad pump linewidth. However, we managed to obtain even smaller Schmidt numbers by optimizing the coupling parameters and spectral width of the pump pulse. In particular, for the ratio of $\kappa_p/\kappa_{is}=6.6$ and the optimal spectral width of $\Delta\omega_{1/2}=0.45\kappa_p$, we have $K=1.0003$. A further increase in $\kappa_p/\kappa_{is}$ to 10 provides $K = 1.00006$, which corresponds to the purity of heralded photons of $\gamma=0.9999$.

\section{Conclusion}
We have shown that a system of optimally coupled ring microresonators is capable of producing almost factorable joint spectral amplitude of the biphoton field, thereby generating near pure heralded single-photon states via spontaneous four-wave mixing. By optimizing the coupling parameters of the system, we present a way for suppressing negative dispersion effects, which, in combination with the optimal spectral width of the pump pulse, provides highest possible purity of the heralded photons generated in such a scheme. The use of resonant coupling via microrings makes it possible to load and unload only required field modes, which may simplify implementation of integrated sources of indistinguishable single photons.

\ack
The work was partially supported by the Russian Science Foundation (project No. 16-12-00045). The results of Sec.~4 were obtained within the state assignment theme No.~0217-2018-0005.

\section*{References}

\bibliographystyle{iopart-num}
\bibliography{WGM_FWM}

\end{document}